\documentstyle[12pt]{article}

\def\be{\begin{equation}}
\def\ee{\end{equation}}
\def\bea{\begin{eqnarray}}    
\def\eea{\end{eqnarray}}

\begin{document}
\setcounter{page}{0}

\begin{flushright}
IPM-97-230  \\
hep-th/9709037
\end{flushright}

\pagestyle{plain}
\begin{center}
\Large {\bf Regularised Supermembrane Theory and Static Configurations 
of M-Theory}\\
\small
(Static Matrix Model)
\vskip .15in

Farhad Ardalan$_{a,b}$, Amir H. Fatollahi$_{a,b}$, 
Kamran Kaviani$_{a,c}$, Shahrokh Parvizi$_{a}$,

\vspace{.5 cm}
\small
{\it a)Institute for Studies in Theoretical Physics and Mathematics (IPM),}\\ 
{\it P.O.Box 19395-5531, Tehran, Iran}\\

\vspace{.3 cm}

{\it b)Department of Physics, Sharif University of Technology,\\
P.O.Box 11365-9161, Tehran, Iran}\\
\vspace{.3 cm}

{\it c)Department of Physics, Az-zahra University,\\
P.O.Box 19834, Tehran, Iran}\\
\vspace{.3 cm}
{\sl E-mails: ardalan, fath, kaviani, parvizi@theory.ipm.ac.ir}

\vskip .2 in
\begin{abstract}
We suggest that the static configurations of M-theory may be described by the
matrix regularisation of the supermembrane theory in static regime. We
compute the long range interaction between an M2-brane and an anti-M2-brane
in agreement with the 11 dimensional supergravity result. 
\end{abstract}
\end{center}


\setcounter{page}{1} 
\pagestyle{plain} 

\section{Introduction} 
The proposed M(atrix) model \cite{BFSS} as a non-perturbative
formulation of M-Theory \cite{To} has provided a new and effective
framework for studying dualities and connections between different
string theories \cite{SS,M,BS,HW,BR}. This model is the dimensional
reduction of 9+1 $U(N)$ SYM theory to 0+1 dimension \cite{HALP} in large
$N$ limit, which latter was introduced and studied as the dynamics of
$N$ D0-branes \cite{W,KPSDF},

In the initial developments of the supermembrane theory
\cite{BST,BST1} (in the 11 dimensional supergravity background) it was
observed that the existence of $\kappa$-symmetry imposes restrictions
on the background fields which reduce to the 11 dimensional
supergravity field equations. Since M-theory has the 11 dimensional
supergravity as its low energy limit, the above observation suggests
that every definition of M-theory should be in close connection to
supermembrane theory. Thus, M-theory in infinite momentum frame and
supermembrane action in light cone gauge written in a matrix form are
related \cite{BFSS}. 

On the other hand the notion of a sub-structure in the formulation of
the M(atrix) model for M-theory has played a central role. Therefore
it is plausible to expect that the same sub-structure in the form of a
matrix formulation should play a role in the framework describing the
static configurations of the M-theory. 

As there is no definition for covariant M-theory, it is tempting to study 
it in various gauges: light-cone, static, etc. . The above 
mentioned relations between supermembrane theory and M-theory in 
light-cone gauge motivates us to search for the similar relation in static 
gauge. Our starting point is the action of supermembranes in 11
dimensions. By restricting the action to the static part of its phase
space we obtain an action which after fixing its $\kappa$-symmetry can
be written in a matrix form. 

The resulted matrix action is invariant under $SO(9)$ rotations of 
target space.
Also the action has a gauge symmetry which corresponds to the
world volume area preserving symmetry. Despite of
the existence of the gauge symmetry, the interpretation of the model as a 
dimensional reduction of a SYM theory seems impossible.

We introduce solutions of the action, which as is expected from M-theory, 
have vanishing quantum corrections. Also we calculate the long
range interaction of parallel M2-brane and anti-M2-brane
solutions of the matrix model. The result is $W(r)\sim \frac{1}{r^6}$ 
which agrees with the uncompactified 11 dimensional supergravity directly
in contrast to the light-cone M(atrix) theory result in compactified 
limit $W(r)\sim \frac{1}{r^5}$.

Conventions and some calculations are gathered in appendices. 
\section{Static supermembrane action as a matrix model}
We use the following notations everywhere:
$$
a,b=0,1,2;\;\;
\mu ,\nu = 0,1,...,9,10;\;\;
I,J,K=1,2,...,9,10;       \;\;
i,j,k=1,2,...,9.
$$
                
The supermembrane action in 11 dimensions is
\cite{DHN,BST}
\bea\label{1}
S={-1\over{2}} \int d^3\eta \bigg
(2\sqrt{-g}
+  \epsilon^{abc} \bar\theta 
\Gamma_{\mu\nu} \partial_a\theta \times
(\Pi_b^\mu \partial_c X^\nu
+{1\over 3} \bar\theta\Gamma^\mu\partial_b\theta\;
\bar\theta\Gamma^\nu\partial_c\theta)\bigg),
\eea
where $\Pi$'s and $g$ are
\bea\label{5}
\Pi_a^\mu= \partial_a X^\mu + \bar\theta 
\Gamma^\mu \partial_a \theta,\;\;\;\;\; g_{ab}=\Pi_a \cdot \Pi_b,
\eea
and $\theta$ is eleven dimensional Majorana spinor. 

The action (\ref{1}) is invariant under global SUSY transformation
\bea\label{10} 
\delta X^{\mu}=-\bar\epsilon\Gamma^{\mu}\theta,\;\;\;
\delta\theta=\epsilon,
\eea
and also under the local fermionic symmetry, $\kappa$-symmetry 
\be \label{15}
\delta X^{\mu}=\bar\kappa(1-\Gamma)\Gamma^{\mu}\theta,\;\;\;\;\;
\delta\theta=(1-\Gamma)\kappa,
\ee
where 
$$
\Gamma={\epsilon^{abc}\over 6\sqrt{-g}} {\Pi_a}^\mu{\Pi_b}^\nu
{\Pi_c}^\rho \Gamma_{\mu\nu\rho},\;\;\;\;\Gamma^2=1.
$$
 
We decompose the coordinates as $\eta_a=(\tau, \sigma_r)$ 
, $r=1,2$.

We go to the static regime defined by
\bea\label{20}
X^0\equiv \tau,     \;\;\;\;
{\dot X}^I \equiv  \dot\theta \equiv 0    ;    
\eea
then the components of $g$ are found to be
\bea\label{25} 
g_{00}&=&-1, \;\;\;\;\;\;\;\;
-f_r\equiv g_{0r}=-\bar\theta\Gamma^0\partial_r\theta, \nonumber\\
g_{rs}&=&\bar{g}_{rs}-f_rf_s,\;\;\;\;\;\;
\bar{g}_{rs}\equiv {\Pi_r}_I {\Pi_s}_I;
\eea
and it can easily be shown that,
\bea\label{30}
g=-\bar{g},\;\;
\bar{g}=det\bar{g}_{rs}={1\over 2} \epsilon^{rs}\epsilon^{r's'} 
\bar{g}_{rr'}\bar{g}_{ss'}=
{1\over 2} (\epsilon^{rs}  \Pi_r^I \Pi_s^J)^2.
\eea

Putting all the above relations in (\ref{1}), we obtain
\bea\label{35} 
S= {1\over{2}} \int d\tau\;d^2\sigma \bigg( -e^{-1} - e\;\bar{g}
-2 {\epsilon}^{rs} \bar\theta\Gamma_{0I}\partial_r\theta \partial_s X^I 
\;- {\epsilon}^{rs} \bar\theta\Gamma_{0I}\partial_r\theta \;
\bar\theta\Gamma^I\partial_s\theta \bigg),
\eea 
where $e$ appears as an auxiliary field for linearising 
the action; its equation of motion gives
\bea\label{37} 
e^2 \bar{g} \;=1,
\eea
which can be used for eliminating $e$. Due to (\ref{37}),
configurations with $\bar{g}=0$ are unacceptable.

The action (\ref{1}) has a local fermionic symmetry, called $\kappa$-symmetry which 
allows one to gauge away half of the fermionic degrees of freedom of 
$\theta$. $\theta$ is a 32-component   11-dimensional 
Majorana spinor and is real 
in a real representation of $\Gamma$ matrices which we use (see appendix).
We fix the $\kappa$-symmetry just as by the light cone gauge
\footnote {In fact we could do gauge fixing before 
restricting the action to its static regime by the ansatz 
(\ref{20}). }
(i.e. $(\Gamma^0+\Gamma^{10})\theta=\Gamma^+\theta=0$) 
\bea\label{40} 
\theta={1\over 2}\left( \begin{array}{l} \lambda \\ \lambda 
\end{array} \right), \;\;\;\;\;\lambda=\lambda^*;
\eea
then it can be shown that
\bea\label{45} 
\bar\theta\Gamma_i\partial\theta&=&0,\;\;\;\;\;\;\;\;
\bar\theta\Gamma_{10}\partial\theta=-{1 \over{2}}
\lambda^T\partial\lambda,\nonumber\\
\bar\theta\Gamma_{0i}\partial\theta&=&-{1 \over{2}} 
\lambda^T\gamma_i\partial\lambda,     \;\;\;\;\;
\bar\theta\Gamma_{0,10}\partial\theta= 0.
\eea
After integration over $\tau$ ( which gives ${\cal{T}}$) 
the action (\ref{35}) takes the following form
\be \label{50}
S=-{1\over {2}}{\cal{T}} \int d^2\sigma e^{-1}
\bigg(\;{1 \over 2} 
{\{X^i,X^j\}}^2+(\{X^i,X^{10}\}-{1\over 2}\lambda^T\{X^i,\lambda\})^2
+\lambda^T\gamma_i\{X^i,\lambda\}+1
\bigg),
\ee 
where  
\be \label{55}
\{ a,b \} =e\;(\partial_{{\sigma}_1}a\partial_{{\sigma}_2}b
-\partial_{{\sigma}_2}a\partial_{{\sigma}_1}b)=
e\;\epsilon^{rs} \partial_ra
\partial_sb,
\ee
which satisfies the Jacobi identity.

We can now formulate our matrix model.
By usual substitutions\cite{DHN,BFSS,IKKT}
\footnote{There is a factor $n$ for $n\times n$ matrices in going 
from bracket to commutator and also from integration to trace. 
Here we absorbed the factor every time in commutator entries.} 
\bea\label{60} 
\{a,b\}\Rightarrow -i\; [ a,b ] ,\;\;\;\;\;
\int \;e^{-1}\; d^2 \sigma\Rightarrow Tr,
\eea 
with the following consequences
\bea\label{65} 
\int \;e^{-1}\; d^2 \sigma \bigg( \{a,b\}c\bigg)=
\int \;e^{-1}\; d^2 \sigma\bigg( a\{b,c\}\bigg) &\Rightarrow&
Tr\bigg([a,b]c\bigg)=Tr\bigg(a[b,c]\bigg) ,
\nonumber\\
\int \;e^{-1}\; d^2 \sigma \{a,b\}=0  &\Rightarrow& Tr[a,b]=0,
\eea 
one finds 
\bea \label{70}
S=&-&  {1\over {2 }}\alpha{\cal{T}}\; Tr \;\bigg(\;{1 \over 2} \;
[X^i,X^j]^2+([X^i,X^{10}]-\gamma {1\over 2}\lambda^T[X^i,\lambda])^2
\;+\;i\lambda^T\gamma_i[X^i,\lambda]\bigg) \nonumber\\
&+& {1 \over 2}\beta {\cal{T}}\; Tr \;(1).
\eea 
Here $\alpha$, $\beta$ and $\gamma$ appeared due to dimensional 
considerations in going from the bracket to the commutator 
and also from integration to trace. We will fix $\alpha$ and $\beta$ later.

The action (\ref{70}) has a gauge symmetry which may be identified with 
area-preserving symmetry of the supermembrane \cite{DHN}. It is defined 
by an arbitrary matrix $\Lambda$
\bea\label{75} 
\delta_{gauge}X^i&=&i[X^i,\Lambda],        \nonumber\\
\delta_{gauge}\lambda&=&i[\lambda,\Lambda],      \nonumber\\  
\delta_{gauge}X^{10}&=&i[X^{10},\Lambda].      
\eea 

Furthermore the action (\ref{70}) is invariant under SUSY 
transformations 
\bea\label{80}
\delta X^i&=& 0,  \nonumber\\    
\delta X^{10}&=&{1\over{2}} \eta^T\lambda,\nonumber\\ 
\delta\lambda&=&\eta,
\eea 
with $\eta$ as an anti-commuting $SO(9)$ spinor and 
it can be shown that the above transformations form space-time SUSY 
algebra
\bea\label{82}
{[}\delta_{\eta},\delta_{\eta'} {]}X^i &=&0,\nonumber\\
{[}\delta_{\eta},\delta_{\eta'} {]}X^{10}&=&\eta'^T\eta,
\nonumber\\
{[}\delta_{\eta},\delta_{\eta'} {]} \lambda&=&0,
\eea
which for $X^{10}$ can be understood as a non-zero translation, due to
$\{ q_A,q_B\}=\Gamma^{10} P_{10}\delta_{AB}$. 
Here 10-th direction is appearing
as the 11-th direction in the super-Galilean algebra \cite{BFSS,BSS}
\footnote{In general to find the complete SUSY transformations, 
 one must search those which respect $\kappa$-symmetry gauge 
fixing solving the equation
$$
\Gamma^+\theta=0 \leftrightarrow \Gamma^+
(\theta+\epsilon+(1-\Gamma)\kappa)=0
\Rightarrow \Gamma^+(\epsilon+(1-\Gamma)\kappa)=0.
$$ 
This is a constraint equation between
SUSY and $\kappa$-symmetry parameters $\epsilon$ and $\kappa$
 as a global and local spinors respectively. 
 A rapid solution is $\kappa=0$ and 
$\epsilon\sim \left( \begin{array}{l} \eta \\ \eta\end{array} \right)$, 
which leads to SUSY transformations (\ref{80}). 
Another closed form solutions seemed unaccessible in our static case. 
A similar observation is reported 
as a result of non-linearities of equations of motion \cite{BST1}. 
So we just keep (\ref{80}).}.
\section{Solutions with vanishing quantum corrections} 
In this section we describe certain configurations which are the
solutions of the classical equations of motion, and it will be shown
that the quantum corrections at one-loop order vanish for them. So these
solutions, as is expected from similar ones in M-theory, show BPS behaviour.

The one-loop effective action around the classical solutions 
$$
X^{10}=\lambda=0, 
$$
is computed in the appendix and the result is
\bea\label{90}       
 \;W= \;{1\over 2}Trlog\bigg(P_i^2\delta_{IJ}-2iF_{ij}\bigg)-
{1 \over 4}\;Trlog\bigg( P_i^2+{i\over{2}}\;
F_{ij}\gamma^{ij}\bigg)-Trlog(P_i^2),
\eea       
with the following definitions
\bea\label{95}       
P_i\;*=[p_i,*],\;\;\;F_{ij}\;*=[f_{ij},*],\;\;\;f_{ij}=i[p_i,p_j],
\eea       
where $p_i$ is classical solution of $X_i$.

Every solution with 
\be\label{100}
F_{ij}=0,\;\;\; \forall{i,j}, 
\ee
leads to vanishing of the one-loop effective action, due 
to the following algebra 
$$
W\sim ({10\over{2}} -{16\over{4}}-1) \;Trlog(P_i^2)\;=0.
$$
In the following we search for these solutions
\footnote
{The point-like configurations which may be represented by 
the following solutions 
$$ 
X^i = diag(x^i_1, x^i_2,...,  x^i_n),\;\;\;
X^{10} = \lambda=0,
$$
are not acceptable because of vanishing $\bar{g}$ in 
(\ref{37}). It is in agreement with the fact that the individual 
11 dimensional supergravitons which are candidates for "quark" 
substructure of our model (due to their role in infinite 
momentum frame  M(atrix) model as "partons") can not be studied as 
static configurations in 11 dimensions, because they are massless.
This argument also will be supported by the equation of motion of $n$, 
the size of matrices. By inserting solutions introduced above,
in the action one finds,
$$
S=0\;+ {1 \over 2}\beta {\cal{T}} n.
$$
The equation of motion for $n$ has no solution (gives $1=0$). }
.

To begin with we consider a solution of (\ref{50}) which
represents a single flat static membrane. With the
conditions $X^{10}=\lambda=0$, the equations of motion (\ref{50}) are
$$
\{X^i,\{X^i,X^j\}\}=0.
$$
Then
\bea\label{102}
X^1=\sigma_1,\;\;\;\;
X^2=\sigma_2,\;\;\;\;
{\rm other}\;\;X^i\;'s=0,
\eea
represent a single membrane solution, $\{X^1,X^2\}=\{\sigma_1,\sigma_2\}=e$
(=1, due to the equation of motion of $e$). In the matrix version 
the conditions $X^{10}=\lambda=0$ give
$$
[X^i,[X^i,X^j]]=0,
$$
which in analogy with (\ref{102}) leads to 
\bea\label{105} 
X^1={L_1\over{\sqrt{2\pi n}}} q,\;\;\;\; 
X^2={L_2\over{\sqrt{2\pi n}}} p, \;\;\;\;
{\rm other}\;\;X^i\;'s=,
\eea 
with $[q,p]=i$ and $0 \leq q,p \leq \sqrt{2\pi n}$ 
eigenvalue distributions. This solution represents a 2 dimensional 
object extended in $X^1$ and $X^2$ directions, and clearly it satisfies 
(\ref{100}) and so is stable under quantum fluctuations. Also due to 
the spectrum of $p$ and $q$ the area of the 2 dimensional object 
(M2-brane) is $L_1 L_2$.

There are also solutions corresponding to two parallel M-2-branes,
\bea\label{110} 
X^1&=&\left( \matrix{
{L_1\over{\sqrt{2\pi n}}} q & 0 \cr
0 & {L_1\over{\sqrt{2\pi n}}} q}\right)\equiv p^1, \;\;\;
X^2=\left( \matrix{
{L_2\over{\sqrt{2\pi n}}} p & 0 \cr
0 & {L_2\over{\sqrt{2\pi n}}} p}\right)\equiv p^2, \nonumber\\
X^3&=&\left( \matrix{
r/2 & 0 \cr
0 & -r/2 }\right)\equiv p^3, \;\;\;\;\;
{\rm other}\;\;X^i\;'s=0, 
\eea
extending in $X^1$ and $X^2$ directions and at the distance 
$r$ in $X^3$ direction. Again clearly this solution satisfies 
(\ref{100}) which means that the two M2-branes are under no-force 
condition. 
\section{M2-brane and anti-M2-brane long range interaction}
In this section we calculate the long range interaction between two parallel
M2-brane and anti-M2-brane.
Solutions corresponding to two membranes with opposite 
charges were introduced in \cite{AB}
\bea\label{115} 
X^1&=&\left( \matrix{
{L_1\over{\sqrt{2\pi n}}} q & 0 \cr
0 & {L_1\over{\sqrt{2\pi n}}} q}\right)\equiv p^1, \;\;\;
X^2=\left( \matrix{
{L_2\over{\sqrt{2\pi n}}} p & 0 \cr
0 & -{L_2\over{\sqrt{2\pi n}}} p}\right)\equiv p^2, \nonumber\\
X^3&=&\left( \matrix{
r/2 & 0 \cr
0 & -r/2 }\right)\equiv p^3, \;\;\;
{\rm other}\;\;X^i\;'s=X^{10}=\lambda=0, 
\eea 
with, $[q,p]=i$ . To calculate the potential 
between these membranes one must find the one-loop effective action of 
(\ref{70}). The one-loop effective action $W$ was introduced 
in the previous section (and calculated in the appendix) 
\be \label{120}       
 \;W= \;{1\over 2}Trlog\bigg(P_i^2\delta_{IJ}-2iF_{ij}\bigg)-
{1 \over 4}\;Trlog\bigg( P_i^2+{i\over{2}}\;
F_{ij}\gamma^{ij}\bigg)-Trlog(P_i^2),
\ee       
with $P_i\;*=[p_i,*]$, $F_{ij}\;*=[f_{ij},*]$, 
$f_{ij}=i[p_i,p_j]$.

The calculation of (\ref{120}) with solutions like (\ref{115}) 
are similar to those 
of \cite{IKKT} for calculating the interaction between
two anti-parallel D-strings.
For solutions (\ref{115}) we have $[p_i,f_{ij}]=c-number$ 
which means that $P_i^2$ and $F_{ij}$ are simultaneously 
diagonalisable. Also $[P_1,P_2]\;\sim i$, which means that  
$P_i^2$ behaves like a harmonic oscillator. 
The steps of calculations are presented in \cite{IKKT} 
and the result is as follows
\be \label{125}
W= (-8n) ({{L_1L_2\over 2\pi n}})^3 {1\over r^6}
+ O({{1\over r^8}}),
\ee
which is in agreement with 11 dimensional supergravity results
for interaction of M2-brane and anti-M2-brane \cite{AB,DS}. It
is notable that this result is in the uncompactified limit
of 11 dimensional supergravity, in contrast 
to the result of light-cone M(atrix) theory ($W(r)\sim {1\over r^5}$)  
\cite{AB}.

The result (\ref{125}) can be used for fixing the parameters 
$\alpha$ and $\beta$ in (\ref{70}). 
By inserting (\ref{105}) in (\ref{70}) one finds
\bea\label{126}
S=({1\over 4})\alpha {\cal{T}} 
({L_1 L_2\over {2\pi n}})^2 \; n + {\beta{\cal{T}}\over 2}n,
\eea
and the equation of motion of $n$ gives 
\bea\label{127}
{L_1 L_2\over {2\pi n}}=\sqrt{{2\beta\over \alpha}},
\eea
resulting in 
\bea\label{128}
S={1\over {2\pi}} \sqrt{{\alpha\beta\over 2}}({\cal{T}}L_1 L_2)
 = T_M  ({\cal{T}}L_1 L_2) .
\eea
which the second equality is the action of a 
flat membrane with $T_M$ as its tension. (\ref{128}) gives 
\bea\label{129}
T_M= {1\over {2\pi}} \sqrt{{\alpha\beta\over 2}}.
\eea

Also by comparing (\ref{125}) with 11 dimensional
supergravity  interaction \cite{AB} one finds
\bea\label{129.1}
{L_1 L_2\over {2\pi n}}=\sqrt{{24\pi {\cal{T}}\over T_M}}.
\eea
By using (\ref{127},\ref{129},\ref{129.1}) and extracting an
irrelevant numerical factor, $\alpha$ and $\beta$ are fixed as follows
\bea\label{129.2}
\alpha =\sqrt{T_M^3\over{\cal{T}}},\;\;\;\;
\beta= 12\pi \sqrt{T_M {\cal{T}}}.
\eea
By choosing ${\cal{T}}=T_M^{-1/3}$ the action (\ref{70}) becomes
\bea \label{129.3}
S=&-&  {1\over {2 }}T_M^{4/3}\; Tr \;\bigg(\;{1 \over 2} \;
[X^i,X^j]^2+([X^i,X^{10}]-\gamma{1\over 2}\lambda^T[X^i,\lambda])^2
\;+\;i\lambda^T\gamma_i[X^i,\lambda]\bigg) \nonumber\\
&+& 6\pi \; Tr \;(1).
\eea 
\section{Conclusion and discussions}
In this letter we introduced a matrix model 
of static configurations of M-theory. By construction the 
large $n$-limit of the model, at least classically, is equivalent  
with static supermembrane action  after 
$\kappa$-symmetry gauge fixing. 
We calculated the long range interaction of an M2-brane 
and an anti-M2-brane solution in this model and found to be in agreement 
with the 11 dimensional supergravity results.

M-theory is supposed  to reduce to various string 
theories and their compactifications. However a model for static 
configurations of M-theory can not be interpreted exactly as a 
string theory, because there are static configuration 
in string theories which are not static in uncompactified M-theory 
(e.g. non-moving D0-branes in IIA theory which are known as KK modes of
massless supergravitons of 11 dimensional supergravity, and so they 
move with speed of light in 11 dimensions.). Notice that the 
reverse of the above argument is not valid, i.e. static configurations 
in M-theory remain static after compactification. So compactifications 
of the static matrix model is  specially interesting.



{\bf Appendix 1- Conventions and notations}\\
Signatures: $g_{ab}=(-,+,+)$, 
$\eta_{\mu\nu}=(-,+,+,+,+,+,+,+,+,+,+)$,\newline 
$\epsilon^{0rs}=-\epsilon^{rs},
\bar\theta=\theta^\dagger\Gamma_0$,
$[\Gamma^\mu,\Gamma^\nu]_+=2\eta^{\mu\nu},\;\;{\Gamma^\mu}^\dagger=
\Gamma^0\Gamma^\mu\Gamma^0$,
$\Gamma^{\mu\nu}=1/2\;(\Gamma^\mu\Gamma^\nu-\Gamma^\nu\Gamma^\mu),$\newline
$\Gamma^0= \left( \matrix{0 & -1_{16}  \cr
 1_{16} & 0 }\right),\;\;\;
\Gamma^{10}= \left( \matrix{1_{16} & 0 \cr
0 & -1_{16} }\right),\;\;\;
\Gamma^i= \left( \matrix{0 & \gamma^i_{16}  \cr
\gamma^i_{16} & 0 }\right),$\newline
$\Gamma^+=\Gamma^0 + \Gamma^{10},\;\;
{\gamma^i_{16}}^\dagger={\gamma^i_{16}}^*=\gamma^i_{16},\;\;
[\gamma^i,\gamma^j]_+=2\delta^{ij},\;\;
\Gamma^1\Gamma^2...\Gamma^9\Gamma^{10}=\Gamma^0.$

{\bf Appendix 2- One-loop effective action}

The calculation of this part is similar to those of \cite{IKKT}.
In this part we decompose the matrices $X$'s and $\theta$'s to classical
solutions and quantum fluctuations as follows, 
\bea\label{140}
X^i&=&(\;p^i\;)_{class.}+a^i,\nonumber\\
\lambda &=& (\;0\;)_{class.} + \phi, \nonumber\\
X^{10}&=& (\;0\;)_{class.}+a^{10},
\eea
where $(...)_{class.}$ are classical solutions and 
the remainder of RHS's are quantum fluctuations around classical solutions. 
After expanding the action (\ref{70}) up to quadratic
terms in fluctuations and using equations of motion one finds
\bea\label{145} 
\Delta S= - Tr \bigg({1\over 2} \; 
[p_i,a_J]^2\;+\;[p_i,p_j][a_i,a_j]  
 - {1\over 2} [p_i,a_i]^2 +{i\over 2} \phi^T\gamma^i[p_i,\phi]\bigg).
\eea 
We have ghosts, because of the gauge invariance introduced
in the text,
$$
S_{ghost}= -Tr\bigg( 
{1\over 2} [p_i,a_i]^2 + [p_i,b][p_i,c]\bigg).
$$
By introducing the adjoint operators
\bea\label{150}
P_i\;*=[p_i,*],\;\;\;F_{ij}\;*=[f_{ij},*],\;\;\;f_{ij}=i[p_i,p_j],
\eea       
the final form of the action will be as follows
$$
S_2= Tr\bigg({1\over 2}( a_I P_i^2\delta_{IJ} a_J - 
a_i 2iF_{ij} a_j)
-\;{i\over 2} \phi^T\gamma^iP_i\phi+bP_i^2c\bigg).
$$
By inserting $S_2$ in the path integral the one-loop effective 
action is obtained  
\bea\label{155}
W&=& - log \int [da][d\phi][dc][db] e^{-S_2}\nonumber\\
&=& \; {1\over 2}\;Trlog\bigg(P_i^2\delta_{IJ}-2iF_{ij}\bigg)-
{1\over 4}Trlog\bigg( P_i^2+{i\over{2}}\;F_{ij}\gamma^{ij}\bigg)-Trlog(P_i^2).
\eea

\end{document}